\documentclass[aps,prb,twocolumn,amsmath,showpacs,amssymb]{revtex4}

\usepackage{amsmath}
\usepackage{graphicx}
\usepackage{amssymb}

\newcommand{\pdag}{{\phantom{\dagger}}}

\begin{document}

\title{Strongly anisotropic spin response as a signature of the helical regime in Rashba nanowires}

\author{Tobias Meng}
\author{Daniel Loss}
\affiliation{Department of Physics, University of Basel, Klingelbergstrasse 82, CH-4056 Basel, Switzerland}

\begin{abstract}
Rashba nanowires in a magnetic field exhibit a helical regime when the spin-orbit momentum is close to the Fermi momentum, $k_F \approx k_{SO}$. We show that this regime is characterized by a strongly anisotropic electron spin susceptibility, with an exponentially suppressed signal along one direction in spin space, and that there are no low frequency spin fluctuations along this direction. Since the spin response in the gapless regime $k_F \not\approx k_{SO}$ has a power law behavior in all three directions, spin measurements provide a signature of the helical regime that complements spin-insensitive conductance measurements.
\end{abstract}


\maketitle
\section{Introduction}
Helical Luttinger liquids, in which the spin of the electrons is locked to their direction of motion, are a central ingredient for a number of recent theoretical proposals and experiments aiming at the detection of zero energy Majorana bound states.\cite{majorana_review} Besides their emergence as topological edge states\cite{qshe1,qshe2,qshe3,qshe4} and their formation in topological insulator nanowires,\cite{egger_yeyati_ti_nonwires_10, ti_nanowires_helical_exp,ti_nanowires_helical_exp2} (quasi-) helical Luttinger liquids can for instance be engineered by placing Carbon nanotubes in an electric field,\cite{klinovaja_cnt_11} by subjecting a Rashba spin-orbit coupled quantum wire (``Rashba nanowire'') to a magnetic field,\cite{streda_03} or by appropriately coupling the electrons in a quantum wire to a Kondo lattice in the RKKY liquid regime, such as the the nuclear spins in the wire.\cite{braunecker_prl_09, braunecker_prb_09, braunecker_10, meng_2bands_13}

One convenient experimental signature of the helical state is provided by the electric conductance through the wire, which drops from $2\,e^2/h$ to $1\,e^2/h$ when the wire becomes helical.\cite{charis_gaas_soi_gap_10} In a Rashba nanowire, this interesting regime can be reached by tuning the chemical potential into the partial gap around zero momentum, see Fig. 1. Even in the absence of spin-orbit interactions, however, lowering the temperature sufficiently can lead to the formation of a helical electronic state due to a spontaneous ordering of the nuclear spins. \cite{braunecker_prl_09,braunecker_prb_09,meng_2bands_13, scheller_13} In the remainder, we argue that the spin physics of the wire provides an additional and in fact complementary signature of this state, which can be probed by spin fluctuation or spin susceptibility measurements. Thanks to considerable experimental advances, these measurements are now believed to be within reach.\cite{stano_13} Different from the conductance, the spin physics depends not only on the presence of a gap, but also on the spin state of the residual gapless modes. We find that the spin susceptibility and spin fluctuations become strongly anisotropic. The susceptibility is exponentially suppressed along the direction set by the spin-orbit coupling in the wire, and there are no low frequency spin fluctuation along this direction. Related physics has been discussed in the context of the Ruderman-Kittel-Kasuya-Yosida (RKKY) interaction mediated by the edge states of quantum spin Hall samples.\cite{gao_rrky_helical_modes_qsh_09} While the spin SU(2) symmetry is broken from the outset by the spin-orbit interaction and the magnetic field, the exponential suppression of the susceptibility along one direction, present only in the helical regime, is markedly different from the anisotropic power law decay of the spin susceptibility in non-helical Rashba nanowires\cite{rkky_anisotropy_bruno} or Carbon nanotubes.\cite{rkky_anisotropy_klinovaja} Our analysis illustrates that interaction effects are important for the experimental detectability of the anisotropic spin physics, and furthermore quantifies the effect of the modes gapped by the combination of spin-orbit coupling and applied magnetic field, which are absent in ideal helical systems. 

The paper is organized as follows. After defining the model in Sec.~\ref{sec:model}, we first discuss the static electron spin susceptibility in Sec.~\ref{sec:suscept}, and contrast the usual response outside the quasi-helical regime to the strongly anisotropic behavior within this regime. In Sec.~\ref{sec:fluct}, we turn to the dynamic spin response of the system, and specifically address the spin fluctuation spectrum, which is also strongly anisotropic. Our results are finally summarized in Sec.~\ref{sec:conclusion}.

\section{The model}\label{sec:model}
To analyze the spin response in the helical regime, we study an interacting single subband quantum wire with sizable Rashba spin-orbit coupling, such as an InAs or InSb wire, \cite{winkler_book} which is subject to a magnetic field parallel to the wire axis and perpendicular to the direction set by the spin-orbit coupling. This setup is depicted in Fig.~\ref{fig:gapped_bands_ll}$(a)$. Choosing the spin-orbit direction as the spin quantization axis, the system can be modeled by the Hamiltonian

\begin{align}
 H &= \int dx\,\sum_{\nu=\uparrow,\downarrow} c_{\nu}^\dagger(x)\left(\frac{-\partial_x^2}{2m}-\mu\right)c_{\nu}^\pdag(x)\nonumber\\
 &-\int dx\,\sum_{\nu,\nu'}\,\frac{k_{SO}}{m}\,c_{\nu}^\dagger(x)\,\sigma^z_{\nu\nu'}\,(-i\partial_x)c_{\nu'}^\pdag(x)\\
 &+\int dx\,\sum_{\nu,\nu'}\,c_{\nu}^\dagger(x)\,\frac{\boldsymbol{\sigma}_{\nu\nu'}}{2}\,c_{\nu'}^\pdag(x)\,\cdot\boldsymbol{B}\nonumber\\
 &+\int dx\,\int dy\,U(x-y)\,\rho(x)\,\rho(y)~,\nonumber
\end{align}
where $c_{\nu}^\pdag(x)$ annihilates an electron of spin $\nu = \uparrow,\downarrow$ at position $x$, the band mass of the electrons is $m$, the chemical potential is $\mu$, the spin-orbit momentum reads $k_{SO}$, the vector of Pauli matrices is given by $\boldsymbol{\sigma}$, and $\rho(x) = \sum_{\nu} c_{\nu}^\dagger(x) c_{\nu}^\pdag(x)$ is the total density at position $x$. The Coulomb interaction, screened on some length scale larger than the width of the wire, is denoted by $U(x-y)$, while $\boldsymbol{B} = (B, 0, 0)^T$ is the applied static and homogeneous magnetic field. In this setup, the latter field is well-known to induce a gap in the electronic spectrum around zero momentum, see Fig.~\ref{fig:gapped_bands_ll}$(b)$. \cite{charis_gaas_soi_gap_10} If the chemical potential is tuned outside the gap, the quantum wire is in a regular spinful Luttinger liquid regime with four gapless modes. If the chemical potential is placed inside the gap, the remaining gapless modes can for our purpose be viewed as a helical Luttinger liquid (note that for the correct treatment of disorder\cite{braunecker_prb_13} or the calculation of observables such as the electronic spectral density or the optical conductivity,\cite{braunecker_prb_12, schuricht_12}  this approximation is insufficient).

\section{Static spin susceptibility}\label{sec:suscept}
We calculate the spin response of the wire by first performing a gauge transformation on the electron operators that trades the spin-orbit interaction for an oscillation in the magnetic field, \cite{braunecker_10}

\begin{align}\label{eq:gauge_trafo}
 c_{\uparrow}^\pdag(x) = e^{i x k_{SO}} \, c_{\uparrow}'(x) \quad,\quad  c_{\downarrow}^\pdag(x) = e^{-i x k_{SO}} \, c_{\downarrow}'(x)~.
\end{align}
This brings the Hamiltonian to the form

\begin{figure}
\centering
(a)\quad\includegraphics[scale=0.35]{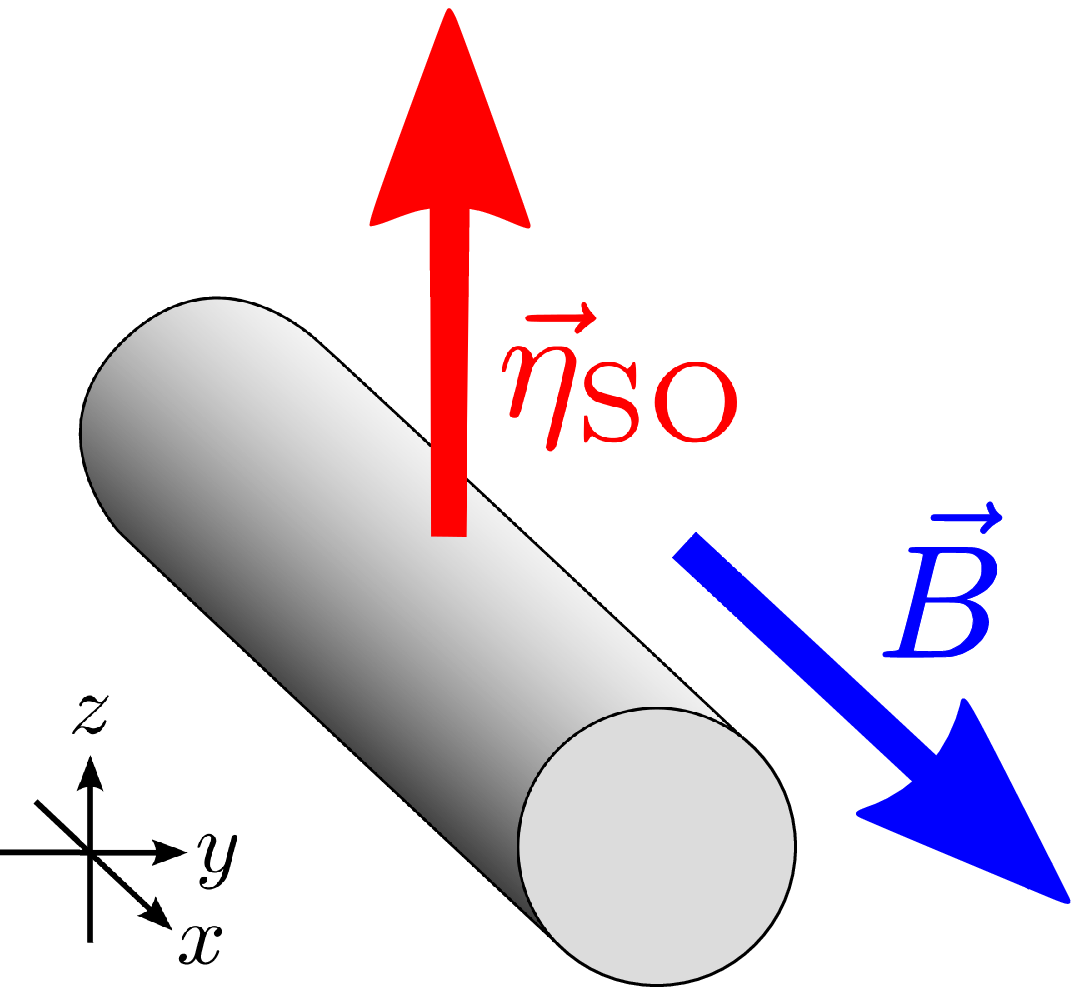}\\[1cm](b)\quad\includegraphics[scale=0.5]{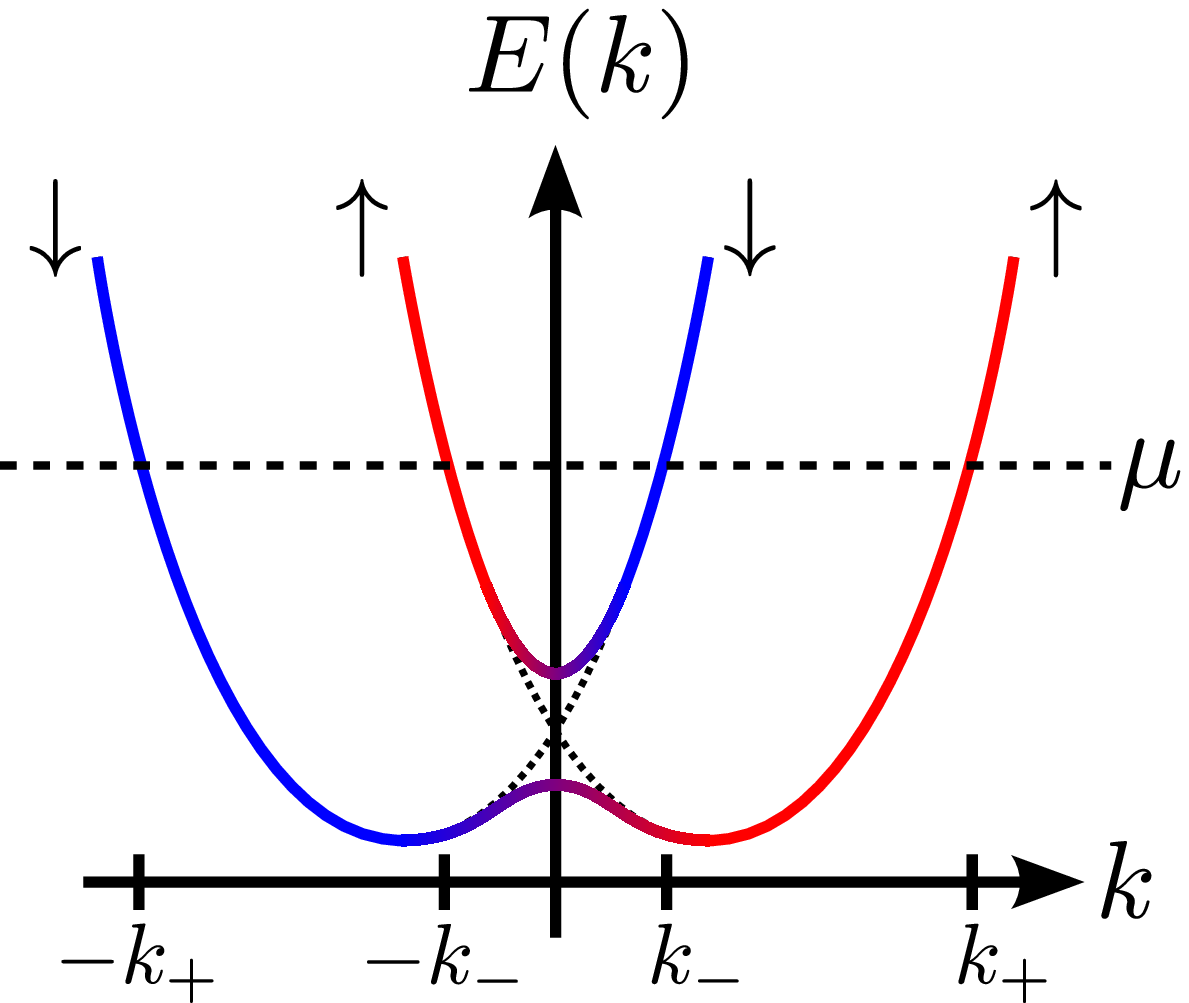}
\caption{Panel $(a)$ depicts the analyzed setup. A quantum wire with Rashba spin-orbit interaction $\sim \boldsymbol{\eta}_{\rm SO}\cdot\boldsymbol{S}$ is subject to a magnetic field parallel to the wire axis and perpendicular to the direction set by the spin-orbit coupling ($\boldsymbol{S}$ denotes the electron spin). Panel $(b)$ shows the spectrum $E(k)$ of the Rashba nanowire as a function of the momentum $k$. The magnetic field $\boldsymbol{B}$ mixes the spin species around $k=0$ and thus opens up a gap. For large chemical potentials $\mu$, the particles close to the Fermi points at $\pm k_+$ and $\pm k_-$ with $k_\pm = k_F \pm k_{SO}$ have spins approximately aligned along the direction set by the spin-orbit interaction. The colors in panel $(b)$ indicates this spin polarization (red corresponds to spin up and blue to spin down). The dotted lines show the spin-polarized bands in the absence of a magnetic field.}
\label{fig:gapped_bands_ll}
\end{figure}

\begin{align}
 H &= \int dx\,\sum_{\nu=\uparrow,\downarrow} c_{\nu}'{}^\dagger(x)\left(\frac{-\partial_x^2}{2m}-\mu- \frac{k_{SO}^2}{2m}\right)c_{\nu}'(x)\label{eq:ham_new_gauge}\\
&+\int dx\,\left(c_{\uparrow}'{}^\dagger(x)c_{\downarrow}'(x)\,e^{-2ixk_{SO}}\,\frac{B}{2} + \text{h.c.}\right)\nonumber\\
 &+\int dx\,\int dy\,U(x-y)\,\rho(x)\,\rho(y)~.\nonumber
\end{align}
After linearizing the spectrum around the Fermi points at momentum $\pm k_F = \pm \sqrt{2m\mu+k_{SO}^2}$, we can treat the wire by standard bosonization techniques.\cite{giamarchi_book} When the chemical potential is tuned far from the gap, such that the system can be viewed as a regular spinful Luttinger liquid, the magnetic field yields only terms that oscillate rapidly at momentum $\pm 2 k_{SO}$ and $\pm 2(k_{SO}\pm k_F)$. For our analysis, these terms can be neglected. The effect of Coulomb interaction, on the other hand, is captured by renormalized Luttinger liquid parameters. The electron spin susceptibility of the wire is now obtained from the imaginary time expression $\chi_{ij}{}'(x-x',\tau-\tau') = \langle T_\tau S_i'(x,\tau)\,S_j'(x',\tau')\rangle$, where $\boldsymbol{S}'(x,\tau) = \sum_{\nu,\nu'} c'_\nu{}^\dagger(x,\tau) \,(\boldsymbol{\sigma}_{\nu\nu'}/2)\,c_{\nu'}'(x,\tau)$ is the electron spin at position $x$ and imaginary time $\tau$. The spin susceptibility in the gapless Luttinger liquid regime has essentially already been derived in Refs.~[\onlinecite{braunecker_prl_09,braunecker_prb_09}]. The experimentally most important static part of the retarded spin susceptibility, which follows from the analytic continuation of the imaginary time expression, diverges at momentum $\pm2k_F$ due backscattering processes,

\begin{subequations}\label{eq:suscept_no_gap}
\begin{align}
 \chi_{xx,yy}^{\rm R}{}'(q,\omega\to0) &\sim \sum_{\kappa=\pm}|q+\kappa\, 2k_F|^{2g_{xy}-2}~,\\
 \chi_{zz}^{\rm R}{}'(q,\omega\to0) & \sim \sum_{\kappa=\pm}|q+\kappa\, 2k_F|^{2g_{z}-2}~,
\end{align}
\end{subequations}
where $q$ denotes the momentum and $\omega$ the frequency, and where $2g_{xy} = K_c + 1/K_s$ and $2g_{z} = K_c + K_s$ are determined by the Luttinger liquid parameters in the charge sector, $K_c$, and in the spin sector, $K_s$. At finite temperatures, the divergences turn into sharp dips. \cite{braunecker_prl_09,braunecker_prb_09} Most importantly, the electron spins thus have a singular response in all three directions. The experimentally measurable spin susceptibilities can be obtained from Eq.~\eqref{eq:suscept_no_gap} by undoing the gauge transformation given in Eq.~\eqref{eq:gauge_trafo}. As a result, the susceptibility in the initial laboratory gauge has components $\chi_{xx} = \chi_{yy}$ with momentum shifted divergences at $q = \pm 2 (k_F+k_{SO})$ and $q = \pm 2 (k_F-k_{SO})$ as compared to gauge-transformed expressions $\chi_{xx}{}' = \chi_{yy}{}'$, while the off-diagonal spin susceptibility $\chi_{xy}$ becomes nonzero in the laboratory gauge, and also diverges at $q = \pm 2 (k_F+ k_{SO})$ and $q = \pm 2 (k_F-k_{SO})$. The form of $\chi_{zz} = \chi_{zz}{}'$, on the other hand, is unchanged, and $\chi_{xz,yz}$ remain zero. The momenta of the divergences of the spin susceptibility correspond to the various possible backscattering processes in the laboratory gauge, as can be inferred from the spectrum shown in Fig.~\ref{fig:gapped_bands_ll}$(b)$.

When the chemical potential is tuned inside the gap, the spin susceptibility should be qualitatively different from Eq.~\eqref{eq:suscept_no_gap}. Fig.~\ref{fig:gapped_bands_ll} indicates that in the helical regime, low energy backscattering is only possible between the two outer Fermi points, and therefore must involve a spin flip. As a consequence, we expect $\chi_{xx,yy,xy}$ are still singular, but only at momentum $q = \pm2(k_{F}+k_{SO})$, while $\chi_{zz}$ should be strongly suppressed because it relies on backscattering processes without spin flip. The electron spin response of the quantum wire should thus be strongly anisotropic (effectively two-dimensional in spin space) when the chemical potential is tuned inside the gap.

To quantify this qualitative argument, we repeat the above analysis for $k_F = k_{SO}$. Starting from Eq.~\eqref{eq:ham_new_gauge}, we decompose the electronic operators into right- and left-movers according to $c_{\nu}'(x) = e^{i xk_F} R_{\nu}(x) + e^{-i xk_F} L_{\nu}(x)$. The latter can be bosonized as $r_{\nu}(z) =(U_{r\nu}/\sqrt{2\pi\alpha})\,e^{-i(r\phi_{\nu}(z)-\theta_{\nu}(z))}$, where $r=R,L\equiv+,-$, while the corresponding Klein factors are denoted as $U_{r\nu}$, and $\alpha$ is a short distance cutoff. \cite{giamarchi_book} Importantly for our discussion, the bosonic fields $\phi_{\nu}$ and $\theta_{\nu}$ are canonically conjugate to each other. As a main difference from the gapless Luttinger liquid regime analyzed above, the magnetic field now yields non-oscillatory cosine potentials for terms connecting left-moving spin up particles and right-moving spin down particles. Introducing the usual spin and charge degrees of freedom via the canonical transformation $\phi_{\hspace*{-2.5pt}\begin{array}{c}\\[-19.5pt] {}_{c}\\[-7pt] {}_{s}\end{array}}(z) = (\phi_{\uparrow}\pm\phi_{\downarrow})/\sqrt{2}$ and $\theta_{\hspace*{-2.5pt}\begin{array}{c}\\[-19.5pt] {}_{c}\\[-7pt] {}_{s}\end{array}}(z) = (\theta_{\uparrow}\pm\theta_{\downarrow})/\sqrt{2}$, and dropping the Klein factors which are not important for our discussion, the non-oscillatory part of the Hamiltonian can be recast into the form

\begin{align}\label{eq:hamiltonian_gapped_regime}
 H&=\int \frac{dx}{2\pi}\,\sum_{i=c,s}\left(\frac{u_i}{K_i} (\partial_x \phi_i)^2 + u_i K_i (\partial_x \theta_i)^2\right)\\
 &+\int dx\,\frac{B}{2\pi\alpha}\,\cos\left(\sqrt{2}(\phi_c+\theta_s)\right)~.\nonumber
\end{align}
As before, $K_c$ and $K_s$ denote the Luttinger liquid parameters in the charge and spin sector, while $u_{c}$ and $u_{s}$ are the corresponding effective velocities. Following Refs.~[\onlinecite{braunecker_prl_09,braunecker_prb_09}], we find that the magnetic field is a relevant perturbation in the renormalization group (RG) sense and gaps out the field $\phi_+ \sim \phi_c + \theta_s$ that corresponds to left-moving spin up particles and right-moving spin down particles. This gap is precisely the gap around zero momentum in the laboratory gauge shown in Fig.~\ref{fig:gapped_bands_ll}.\cite{braunecker_10} In order to calculate the electron spin susceptibility in this partially gapped regime, we perform a canonical transformation that switches from the spin and charge degrees of freedom to the field $\phi_{+} \sim \phi_{c}+\theta_{s}$ and an appropriate linearly independent combination of $\phi_c$ and $\theta_s$, \cite{braunecker_prb_09}

\begin{subequations}\label{eq:field_trafo}
\begin{align}
 \phi_{c} &=\frac{K_{c}}{\sqrt{K}}\,\phi_{+} + \sqrt{\frac{K_{c}}{K_{s}\,K}}\,\phi_{-}~,\\
  \theta_{c} &=\frac{1}{\sqrt{K}}\,\theta_{+} + \frac{1}{\sqrt{K_{c}\,K_{s}\,K}}\,\theta_{-}~,\\
  \phi_{s} &=\frac{1}{\sqrt{K}}\,\theta_{+} - \sqrt{\frac{K_{s}\,K_{c}}{K}}\,\theta_{-} ~,\\
  \theta_{s}& = \frac{1}{K_{s}\sqrt{K}}\,\phi_{+} - \sqrt{\frac{K_{c}}{K_{s}\,K}}\,\phi_{-}~,
\end{align}
\end{subequations}
with $K = K_{c} + 1/K_{s}$. The RG equation for the magnetic field may now be derived in a real space RG analysis that parametrizes the running short distance cutoff as $\alpha(b) = \alpha\,b$. It reads \cite{braunecker_prb_09}

\begin{align}
 \frac{\partial B}{\partial \log(b)} = (1-K/2)\,B\text{ .}\label{eq:rg_overhauser}
\end{align}
The magnetic field is thus RG relevant for $K < 2$, which is fulfilled in interacting quantum wires. \cite{auslaender_spin_charge_seperation_exp_05, steinberg_charge_fractionalization_08,jompol_kc_ks} The RG flow is integrated until the length scale $u_+(b)/\Delta(b)$ associated with the running gap $\Delta(b)$ of $\phi_+$ equals the running short distance cutoff (at a given RG step, this gap can be defined by the expansion of the sine-Gordon potential to second order, which is strictly speaking only justified at the end of the flow\cite{giamarchi_book}). We obtain the low energy Hamiltonian at the end of the flow as

\begin{align}
 H &=\int \frac{dx}{2\pi}\, \left(\frac{u_+^*}{K_+^*}(\partial_x \phi_+)^2 + \frac{\Delta^2}{K_+^* u_+^*}\,\phi_+^2 + u_+^* K_+^*(\partial_x \theta_+)^2\right)\nonumber\\
 &+\int \frac{dx}{2\pi}\,\left(\frac{u_-^*}{K_-^*}(\partial_x \phi_-)^2 + u_-^* K_-^*(\partial_x \theta_-)^2\right)\\
 &+\int \frac{dx}{2\pi} \left( U_\phi^*\,(\partial_x \phi_+) (\partial_x \phi_-) +U_\theta^*\,(\partial_x \theta_+) (\partial_x \theta_-)\right) ~,\nonumber
\end{align}
where $u_\pm^*$ and $K_\pm^*$ are the strong coupling values of the velocities and Luttinger liquid parameters in the $\pm$-channel, while the gap is $\Delta = u_+^*/\alpha^*$ with $\alpha^*$ being the renormalized short distance cutoff. The interactions $U_\phi^*$ and $U_\theta^*$ are the strong coupling values of the interactions introduced by the canonical transformation given in Eq.~\eqref{eq:field_trafo}. These interactions constitute further subleading corrections,\cite{braunecker_prb_09} which essentially renormalize the Luttinger liquid parameters and velocities. In a mean field picture, the interaction $U_\phi^*$ is subleading because the field $\phi_+$ is pinned to one of the minima of the sine-Gordon potential in Eq.~\eqref{eq:hamiltonian_gapped_regime}. Fluctuations around the mean field are suppressed by the gap $\Delta$ that is of the order of the bandwidth of the renormalized theory. The interaction $U_\theta^*$ is most conveniently analyzed by switching from the Hamiltonian to the associated (imaginary time $\tau$) action and integrating out $\theta_\pm$. This yields an additional small renormalization of the velocities and Luttinger liquid parameters $u_\pm^*$ and $K_\pm^*$, plus an interaction of the form $(\partial_\tau \phi_+)(\partial_\tau \phi_-)$, which is subleading for the same reason as $U_\phi^*$. We will therefore from now on consider $u_\pm^*$ and $K_\pm^*$ to be renormalized values that also account for the effect of the off-diagonal terms on velocities and Luttinger liquid parameters and drop $U_\phi^*$ and $U_\theta^*$ in the remainder. We furthermore disregard solitons connecting the different minima of the sine-Gordon potential, which alter the properties of the wire at temperatures lower than the typical experimental ones, as well as its finite frequency response. \cite{ponomarenko_98, braunecker_prb_12, schuricht_12}

These considerations finally allow the calculation of the spin susceptibilities in the helical regime. As before, we only keep the backscattering contributions, since forward scattering is non-singular. In the initial laboratory gauge, the $x$ and $y$ components of the imaginary time spin susceptibility read

\begin{align}
& \chi_{xx}(x,\tau) = \chi_{yy}(x,\tau) \\
 &= \frac{1}{4}e^{-i2x(k_F+k_{SO})}\langle T_\tau R_{\uparrow}^\dagger(x,\tau)L_{\downarrow}^\pdag(x,\tau)L_{\downarrow}^\dagger(0,0)R_{\uparrow}^\pdag(0,0)\rangle\nonumber\\
  &+ \frac{1}{4}e^{-i2x(k_F-k_{SO})}\langle T_\tau R_{\downarrow}^\dagger(x,\tau)L_{\uparrow}^\pdag(x,\tau)L_{\uparrow}^\dagger(0,0)R_{\downarrow}^\pdag(0,0)\rangle\nonumber\\
  &+\text{h.c.}\nonumber
\end{align}
Bosonizing these expressions and performing the canonical transformation given in Eq.~\eqref{eq:field_trafo}, we obtain

\begin{align}
& \chi_{xx}(x,\tau) = \chi_{yy}(x,\tau) =\frac{1}{4(2\pi\alpha)^2}e^{-i2x(k_F+k_{SO})}\\
 &\times\langle e^{i\sqrt{2}(K_c-1/K_s)(\phi_+(x,\tau)-\phi_+(0,0))/\sqrt{K}}\rangle\nonumber\\
 &\times \langle e^{i\sqrt{2}\sqrt{4K_c/(K_s K)}(\phi_-(x,\tau)-\phi_-(0,0))}\rangle\nonumber\\
  &+\frac{1}{4(2\pi\alpha)^2}e^{-i2x(k_F-k_{SO})}\,\langle e^{i\sqrt{2}\sqrt{K}(\phi_+(x,\tau)-\phi_+(0,0))}\rangle\nonumber\\
 &+\text{h.c.}\nonumber
\end{align}
In the gapped regime, the field $\phi_+$ is pinned, and the expectation values involving this field can be approximated by 1. One may also go beyond this mean field argument by noting that $\langle e^{i\sqrt{2}A(\phi_+(x,\tau)-\phi_+(0,0))}\rangle =  e^{-A^2\langle(\phi_+(x,\tau)-\phi_+(0,0))^2\rangle}$, and that the correlation function of $\phi_+$ decays exponentially due to the gap. Therefore, the expectation values involving $\phi_+$ are exponentials of an exponential and indeed go to unity very quickly. The remaining average over the gapless field $\phi_-$ yields the usual Luttinger liquid power law decay. The components of the spin susceptibility perpendicular to the spin-orbit axis are thus given by

\begin{align}
&\chi_{xx}(x,\tau) = \chi_{yy}(x,\tau) \approx \label{eq:suscept_gapless} \\
 &\frac{1}{4(2\pi\alpha)^2}e^{-i2x(k_F+k_{SO})}\,\left(\frac{\alpha}{\sqrt{x^2+(u_-^* |\tau|+\alpha)^2}}\right)^{\frac{4K_cK_-^*}{K_s K}}\nonumber\\
&+\frac{1}{4(2\pi\alpha)^2}e^{-i2x(k_F-k_{SO})}+\text{h.c.}\nonumber
\end{align}
When performing the analytical continuation in order to derive the physically relevant retarded spin susceptibility, the oscillating factor in the last line of Eq.~\eqref{eq:suscept_gapless},  stemming from the gapped mode $\phi_+$, drops out, and only a Luttinger liquid power law deriving from the gapless mode remains.\cite{giamarchi_book,braunecker_prb_09} A similar power law is found for $\chi_{xy}(x,\tau)$,

\begin{align}
&\chi_{xy}(x,\tau)  \approx \label{eq:suscept_gapless_xy} \\
 &\frac{i}{4(2\pi\alpha)^2}e^{-i2x(k_F+k_{SO})}\,\left(\frac{\alpha}{\sqrt{x^2+(u_-^* |\tau|+\alpha)^2}}\right)^{\frac{4K_cK_-^*}{K_s K}}\nonumber\\
&-\frac{i}{4(2\pi\alpha)^2}e^{-i2x(k_F-k_{SO})}+\text{h.c.}\nonumber
\end{align}
The mixed susceptibilities $\chi_{xz}$ and $\chi_{yz}$ vanish. The susceptibility along $z$, on the other hand, reads

\begin{align}
& \chi_{zz}(x,\tau) =\frac{1}{4(2\pi\alpha)^2}e^{-i2xk_F}\label{eq:chi_z_1}\\
 &\times\langle e^{i\sqrt{2/K}(K_c\phi_+(x,\tau) + \theta_+(x,\tau)-K_c\phi_+(0,0) - \theta_+(0,0))}\rangle\nonumber\\
 &\times \langle e^{i\sqrt{2 K_c/(K_sK)}(\phi_-(x,\tau) -K_s \theta_-(x,\tau)-\phi_-(0,0) + K_s\theta_-(0,0))}\rangle\nonumber\\
  &+\frac{1}{4(2\pi\alpha)^2}e^{-i2xk_F}\nonumber\\
 &\times\langle e^{i\sqrt{2/K}(K_c\phi_+(x,\tau) - \theta_+(x,\tau)-K_c\phi_+(0,0) + \theta_+(0,0))}\rangle\nonumber\\
 &\times \langle e^{i\sqrt{2 K_c/(K_sK)}(\phi_-(x,\tau) +K_s \theta_-(x,\tau)-\phi_-(0,0) - K_s\theta_-(0,0))}\rangle\nonumber\\
 &+\text{h.c.}\nonumber
\end{align}
Again, the field $\phi_+$ can be replaced by its average value and drops out. The field $\theta_+$, being canonically conjugate to the ordered field $\phi_+$, has large fluctuations that suppress $\chi_{zz}$. As has been established in Refs.~[\onlinecite{voit_98,starykh_99,braunecker_prb_12}], and neglecting the additional phase factor due to the simultaneous presence of $\theta$ and $\phi$ fields, Eq.~\eqref{eq:chi_z_1} can be evaluated as
 
\begin{align} 
 &\chi_{zz}(x,\tau)\approx\frac{1}{2(2\pi\alpha)^2} e^{-i2xk_F}\label{eq:suscept_gapped_z}\\
 &\times\left(\frac{\alpha}{\sqrt{x^2+(u_-^* |\tau|+\alpha)^2}}\right)^{K_c(K_-^*+K_s^2/K_-^*)/(K_s K)}\nonumber\\
 &\times\left(\frac{\alpha}{\sqrt{x^2+(u_+^* |\tau|+\alpha)^2}}\right)^{1/(K_+^*K)}\,e^{-\frac{\mathcal{C}\,\Delta^*}{K_+^*Ku_+^*}\sqrt{x^2 + (u_+^*\tau)^2}}\nonumber\\
 &+\text{h.c.}~,\nonumber
\end{align}
where $\mathcal{C}$ is a constant of order one. The spin susceptibility in $z$ direction is thus indeed exponentially suppressed by the gap. The associated typical length scale is given by the renormalized short distance cut-off of the theory, $u_+^*/\Delta = \alpha^*$. The suppression of the signal along $z$ is increased by electron-electron interactions, which strongly enhance the gap according to Eq.~\eqref{eq:rg_overhauser}. 

The static parts of the retarded spin susceptibility in the momentum/frequency-domain can now be obtained by Fourier transformation and analytic continuation. In $x$ and $y$ direction, this yields the expression

\begin{align}
 \chi_{xx,yy,xy}^{\rm R}(q,\omega\to0) & \sim \sum_{\kappa=\pm}|q+\kappa 2(k_F+k_{SO})|^{\frac{4K_cK_-^*}{K_s K}-2}~.\label{eq:chi_xx_yy}
\end{align}
With the experimental values $K_c \approx 0.5$ and $K_s\approx 1$, \cite{steinberg_charge_fractionalization_08, jompol_kc_ks} we find that $\chi_{xx} = \chi_{yy}$ and $\chi_{xy}$ diverge at zero temperature. Similar to the renormalization of the gap, electron-electron interactions in the wire also strengthen the divergence in Eq.~\eqref{eq:chi_xx_yy} through a decrease of the value of $K_c$. In real space, on the other hand, stronger interactions correspond to a weaker power law decay of the signal along $x$ and $y$ at large distances. 

Along $z$, we use the fact that a Yukawa potential-like function $f(x,y) = (1/\sqrt{x^2+y^2})^n\,e^{-\Delta\sqrt{x^2+y^2}}$ has the Fourier transform 

\begin{align}
f(q_x, q_y) &= \int_0^\infty dr \int_0^{\pi} d\varphi\,e^{i q r \cos(\varphi)}\,\int_{-\infty}^{\infty}\frac{dk}{\pi}\,e^{-i k r} f_\Delta(k)\nonumber\\
&= \int_0^{\pi} \frac{d\varphi}{2} \,f_\Delta(q \cos(\varphi))\label{eq:chi_xy}
\end{align}
with $f_\Delta(k) \sim (1/\sqrt{k^2+\Delta^2})^{2-n}$ (here, we use $y \sim u_{\pm}^* \tau \in [0,\infty]$ at zero temperature and neglect the difference in the effective velocities). The backscattering contribution to the susceptibility in $z$ direction is thus given by a small, non-singular expression proportional to an inverse power of the gap, such that stronger interactions result in a further suppression of the signal. Because also the forward scattering is non-singular,\cite{braunecker_prb_09} the spin susceptibility of a Rashba spin-orbit coupled quantum wire in the helical regime is strongly suppressed along the direction of the spin-orbit coupling. This renders the spin response effectively two-dimensional in spin space.

\section{Dynamic susceptibilities and spin fluctuations}\label{sec:fluct}
The strongly anisotropic character of the susceptibility implies, by virtue of the fluctuation-dissipation theorem, that also the spin fluctuations along the direction set by the spin-orbit coupling are suppressed for frequencies below the gap. To show this, we calculate the spectral function of the spin fluctuations,

\begin{align}
 \mathcal{S}_{\sigma\sigma'}(x,\omega) &= \int_{-\infty}^{\infty} dt\,e^{i\omega t}\,\mathcal{S}_{\sigma\sigma'}(x,t)~,\\
 \mathcal{S}_{\sigma\sigma'}(x,t) &= \frac{1}{2}\,\langle S_\sigma(x,t) S_{\sigma'}(0,0)+S_\sigma(0,0) S_{\sigma'}(x,t) \rangle~.\nonumber
\end{align}
The latter is related to the imaginary part of the corresponding susceptibility

\begin{align}
 \chi_{\sigma\sigma'}^R(x,\omega) &= \int_{-\infty}^{\infty} dt\,e^{i\omega t}\, i\theta(t)\,\langle [S_\sigma(x,t), S_{\sigma'}(0,0)]\rangle
\end{align}
by the the fluctuation-dissipation theorem,

\begin{align}
 \mathcal{S}_{\sigma\sigma'}(x,\omega) = \coth\left(\frac{\beta\omega}{2}\right)\,\mathfrak{Im}\left\{ \chi_{\sigma\sigma'}^R(x,\omega)\right\}~,
\end{align}
where $\beta = T^{-1}$ is the inverse temperature in units of $k_B = 1$. For the $x$ and $y$ directions, the retarded real time spin susceptibilities can be obtained from Eq.~\eqref{eq:suscept_gapless}. They are given by\cite{giamarchi_book}

\begin{align}
&\chi_{xx}^R(x,t) = \chi_{yy}^R(x,t)  \label{eq:suscept_gapless_ret} \\
 &\approx\theta(u_-^*t-|x|)\,\frac{\sin(\pi\tilde{K})\,\cos(2(k_F+k_{SO})x)}{(2\pi\alpha)^2}\nonumber\\
 &\times\left(\frac{\alpha^2}{(u_-^*t)^2-x^2}\right)^{\tilde{K}}\nonumber~,
\end{align}
at zero temperature, where $\tilde{K} = 2K_cK_-^*/(K_sK)$. From Eq.~\eqref{eq:chi_xx_yy}, we find an analogous expression for $\chi_{xy}^R(x,t)$ with $\cos(2(k_F+k_{SO})x)\to\sin(2(k_F+k_{SO})x)$. The imaginary part of the Fourier transform of this expression yields the zero temperature spectrum of the spin fluctuations as

\begin{align}
 &\mathcal{S}_{xx}(x,\omega) = \mathcal{S}_{yy}(x,\omega)\label{eq:fluctuations_xy}\\
 &\approx \frac{\sin(\pi\tilde{K})\,\cos(2(k_F+k_{SO})x)\sqrt{\pi}2^{-\tilde{K}-1/2}\Gamma(1-\tilde{K})}{(2\pi\alpha)^2}\nonumber\\
 &\times \frac{\alpha}{u_-^*}\,\left|\frac{\omega\, \alpha^2}{u_-^*\,x}\right|^{\tilde{K}-1/2}\,J_{\tilde{K}-1/2}\left(\left|\frac{\omega\,x}{u_-^*}\right|\right)\,\text{sgn}(\omega)~,\nonumber
 \end{align}
where $J_\alpha$ is a Bessel function of the first kind and $\Gamma$ is the standard Gamma function. For small frequencies $\omega \ll u_-^*/x$, the spin fluctuation spectrum is thus proportional to $|\omega|^{2\tilde{K}-1}\,\text{sgn}(\omega)$, as could have been expected from a dimensional analysis of Eq.~\eqref{eq:suscept_gapless_ret}. In the $z$ direction, on the other hand, the fluctuations are gapped. This implies a vanishing $\mathcal{S}_{zz}(x,\omega)$ for frequencies $|\omega| < \Delta$, as can be shown by Fourier transformation of $\chi_{zz}(q,\omega_n)$ and subsequent analytic continuation. Like the spin susceptibility, spin fluctuations are thus strongly anisotropic for frequencies smaller than the gap, and again, this anisotropy is strengthened by electron-electron interactions, which increase the gap $\Delta$ and weaken the power law suppression of $\mathcal{S}_{xx}$, $\mathcal{S}_{yy}$, and $\mathcal{S}_{xy}$ and  at low frequencies.

\section{Conclusions}\label{sec:conclusion}
In this work, we showed that a Rashba nanowire in the helical regime (and more generally any helical or quasi-helical Luttinger liquid) exhibits strongly anisotropic spin physics, and analyzed the latter in terms of the static spin susceptibility and the dynamic spin response. Given that Rashba nanowires can be mapped onto quantum wires with helical nuclear spin order,\cite{braunecker_10} the same anisotropic spin physics also provides a specific signature of helical nuclear order in quantum wires. As discussed in Sec.~\ref{sec:suscept}, the helical regime is characterized by an exponentially suppressed static spin susceptibility along the direction set by the spin-orbit coupling, while it shows a power-law decay in the perpendicular directions. Outside the helical regime, on the other hand, the susceptibility exhibits a power law decay along all three directions. A strongly anisotropic behavior was also obtained for the dynamic properties of the spins, as has been discussed in Sec.~\ref{sec:fluct}. In particular, we found that the spin fluctuation spectrum along the direction set by the spin-orbit interaction vanishes for frequencies below the gap, while it behaves as an interaction-dependent frequency power law in the perpendicular directions.  We furthermore discussed that the strongly anisotropic character of the spin physics as well as the detectability of the susceptibility and the fluctuation spectrum in the perpendicular directions are importantly increased by electron-electron interactions.  In conclusion, spin physics provides an additional experimental signature of the helical regime, and complements transport measurement \cite{charis_gaas_soi_gap_10,scheller_13} and possible tunneling spectroscopy experiments. \cite{auslaender_spectroscopy_02} Different from conductance measurements, which give only access to the number of gapless modes, the spin physics depends on the spin state of these modes. A gap for one of the two spin  species would for instance result in a similar reduction of the conductance, but would yield a spin response along a single direction.

\acknowledgements
We would like to thank Peter Stano for helpful discussions. This work has been supported by SNF, NCCR Nano, and NCCR QSIT.



\begin{thebibliography}{99}

\bibitem{majorana_review}
For a review, see for instance J. Alicea, Rep. Prog. Phys. \textbf{75}, 076501 (2012).

\bibitem{qshe1}
C. L. Kane and E. J. Mele, Phys. Rev. Lett. \textbf{95}, 226801 (2005).

\bibitem{qshe2}
B. A. Bernevig and S.-C. Zhang, Phys. Rev. Lett. \textbf{96}, 106802 (2006).

\bibitem{qshe3}
M. K\"{o}nig, S. Wiedmann, C. Br\"{u}ne, A. Roth, H. Buhmann, L. W. Molenkamp, X.-L. Qi, and S.-C. Zhang, Science \textbf{318}, 766 (2007).

\bibitem{qshe4}
M. Z. Hasan and C. L. Kane, Rev. Mod. Phys. 82, \textbf{3045} (2010).

\bibitem{egger_yeyati_ti_nonwires_10}
R. Egger, A. Zazunov, and A. Levy Yeyati, Phys. Rev. Lett. \textbf{105}, 136403 (2010).

\bibitem{ti_nanowires_helical_exp}
D. Kong, J. C. Randel, H. Peng, J. J. Cha, S. Meister, K. Lai, Y. Chen, Z.-X. Shen, H. C. Manoharan, and Y. Cui, Nano Lett. \textbf{10}, 329 (2010).

\bibitem{ti_nanowires_helical_exp2}
H. Peng, K. Lai, D. Kong, S. Meister, Y. Chen, X.-L. Qi, S.-C. Zhang, Z.-X. Shen, and Y. Cui, Nat. Mater. 9, 225 (2010).

\bibitem{klinovaja_cnt_11}
J. Klinovaja, M. J. Schmidt, B. Braunecker, and D. Loss, Phys. Rev. B \textbf{84}, 085452 (2011).

\bibitem{streda_03}
P. St\ifmmode \check{r}\else \v{r}\fi{}eda and P. \ifmmode \check{S}\else \v{S}\fi{}eba, Phys. Rev.  Lett. \textbf{90}, 256601 (2003).

\bibitem{braunecker_10}
B. Braunecker, G. I. Japaridze, J. Klinovaja, and D. Loss, Phys. Rev. B \textbf{82}, 045127 (2010).

\bibitem{braunecker_prl_09}
B. Braunecker, P. Simon, and D. Loss, Phys. Rev. B \textbf{80}, 165119 (2009).


\bibitem{braunecker_prb_09}
B. Braunecker, P. Simon, and D. Loss, Phys. Rev. Lett. \textbf{102}, 116403 (2009).


\bibitem{meng_2bands_13}
T. Meng and D. Loss, Phys. Rev. B \textbf{87}, 235427 (2013).

\bibitem{charis_gaas_soi_gap_10}
C. H. L. Quay, T. L. Hughes, J. A. Sulpizio, L. N. Pfeiffer, K. W. Baldwin, K. W. West, D. Goldhaber-Gordon, and R. de Picciotto, Nature Physics \textbf{6}, 336 (2010). 

\bibitem{scheller_13}
C. P. Scheller, T.-M. Liu, G. Barak, A. Yacoby, L. N. Pfeiffer, K. W. West, and D. M. Zumb\"{u}hl, arXiv:1306.1940.

\bibitem{stano_13}
P. Stano, J. Klinovaja, A. Yacoby, and D. Loss, arXiv:1303.1151.

\bibitem{gao_rrky_helical_modes_qsh_09}
J. Gao, W. Chen, X. C. Xie, and F.-C. Zhang, Phys. Rev. B \textbf{80}, 241302(R) (2009).

\bibitem{rkky_anisotropy_bruno}
H. Imamura, P. Bruno, and Y. Utsumi, Phys. Rev. B \textbf{69}, 121303(R) (2004).

\bibitem{rkky_anisotropy_klinovaja}
J. Klinovaja and D. Loss, Phys. Rev. B \textbf{87}, 045422 (2013).

\bibitem{winkler_book}
R. Winkler, {\it Spin-Orbit Coupling Effects in Two-Dimensional Electron and Hole Systems} (Springer, New York, 2003).

\bibitem{braunecker_prb_13}
B. Braunecker, A. Str\"{o}m, and G. I. Japaridze, Phys. Rev. B \textbf{87}, 075151 (2013).

\bibitem{braunecker_prb_12}
B. Braunecker, C. Bena, and P. Simon,  Phys. Rev. B \textbf{85}, 035136 (2012).

\bibitem{schuricht_12}
D. Schuricht, Phys. Rev. B \textbf{85}, 121101(R) (2012).

\bibitem{giamarchi_book}
T. Giamarchi, {\it Quantum Physics in One Dimension} (Oxford University Press, New York, 2003).

\bibitem{auslaender_spin_charge_seperation_exp_05}
O. M. Auslaender, H. Steinberg, A. Yacoby, Y. Tserkovnyak, B. I. Halperin, K. W. Baldwin, L. N. Pfeiffer, and K. W. West, Science \textbf{308}, 88 (2005).

\bibitem{jompol_kc_ks}
Y. Jompol, C. J. B. Ford, J. P. Griffiths, I. Farrer, G. A. C. Jones, D. Anderson, D. A. Ritchie, T. W. Silk, and A. J. Schofield, Science \textbf{325}, 597 (2009).

\bibitem{steinberg_charge_fractionalization_08}
H. Steinberg, G. Barak, A. Yacoby, L. N. Pfeiffer, K. W. West, B. I. Halperin, and K. Le Hur, Nature Physics \textbf{4}, 116 (2008).

\bibitem{ponomarenko_98}
V. V. Ponomarenko and N. Nagaosa, Phys. Rev. Lett. \textbf{81}, 2304 (1998).

\bibitem{voit_98}
J. Voit, Eur. Phys. J. B \textbf{5}, 505 (1998).

\bibitem{starykh_99}
O. A. Starykh, D. L. Maslov, W. H\"{a}usler, and L. I. Glazman in {\it Low-Dimensional Systems: Interactions and Transport Properties}, Lecture Notes in Physics, edited by T. Brandes (Springer, New York, 2000), Part I, p. 37. (Springer, 2000); or arxiv:cond-mat/9911286.

\bibitem{auslaender_spectroscopy_02}
O. M. Auslaender, A. Yacoby, R. de Picciotto, K. W. Baldwin, L. N. Pfeiffer, and K. W. West, Science \textbf{295}, 825 (2002).

\end{thebibliography}
\end{document}